\documentclass[12pt]{article}
\usepackage{amsfonts,graphicx,amssymb,amsmath,amsthm,epsfig}
\usepackage{float}
\allowdisplaybreaks
\begin{document}
\title{\textbf{New Agegraphic Dark Energy Model in Modified Symmetric Teleparallel Theory}}
\author{Madiha Ajmal \thanks {madihaajmal222@gmail.com}~~and
M. Sharif \thanks {msharif.math@pu.edu.pk} \\
Department of Mathematics and Statistics, The University of Lahore,\\
1-KM Defence Road Lahore-54000, Pakistan.}

\date{}
\maketitle

\begin{abstract}
In this manuscript, we examine the cosmological significance of the
new agegraphic dark energy model by investigating different
cosmological parameters such as the equation of state parameter,
$\omega_{D}-\omega^{\prime}_{D}$ and the $r-s$ planes in the
framework of $f(\mathcal{Q})$ theory. We consider flat
Friedmann-Robertson-Walker universe model under interacting
conditions between dark energy and dark matter. The equation of
state parameter indicates a quintessence-like characteristic of the
universe. The stability of the model is analyzed using the squared
speed of sound parameter which demonstrates the unstable behavior of
the new agegraphic dark energy model throughout the cosmic
evolution. The freezing region is represented by the
$\omega_{D}-\omega^{\prime}_{D}$ plane, while the Chaplygin gas
model corresponds to the $r-s$ plane. It is worthwhile to mention
here that the interacting new agegraphic dark energy model addresses
the cosmic coincidence problem by allowing the energy density ratio
between dark energy and dark matter to evolve slowly over cosmic
time.
\end{abstract}
\textbf{Keywords}: New agegraphic dark energy; $f(\mathcal{Q})$
gravity; Cosmological evolution.\\
\textbf{PACS}: 95.36.+x; 04.50.Kd; 64.30.+t.

\section{Introduction}

The study of large-scale structures, supernova type-Ia and cosmic
microwave background radiations have presented compelling evidences
indicating that our universe is primarily characterized by two
mysterious components, dark matter (DM) and dark energy (DE)
\cite{1}. Dark energy drives the current rapid expansion of the
cosmos, while DM contributes to explain the rotation curves of
galaxies and the overall structure of the universe. In the context
of DE models, the rapid expansion has been discussed by altering the
energy-momentum tensor (EMT) which is directly related to the
right-hand side of the Einstein field equations. The modified theory
of gravity involves altering the geometric aspect on the left-hand
side of the field equations. Therefore, we are still a long way from
creating a complete theory that can explain not only the rapid
expansion of the universe but also problems with early cosmology,
structure development, DM and other difficulties. Regardless of this
approach, it is essential to include quantum effects to develop a
precise theory of gravity. The quantum gravitational theory is the
theory of gravity that includes the ideas of quantum mechanics.
While quantum gravity remains an unresolved theory, several ideas
have been suggested based on its principle. Holographic DE (HDE) and
agegraphic DE (ADE) have been proposed as possible candidates for
explaining the recent accelerated expansion of the universe by
incorporating key properties of quantum gravity. The DE models offer
a comprehensive framework for understanding the universe and solving
various challenges in modern cosmology such as the coincidence
problem \cite{3}.

The ADE framework originates from quantum mechanics based on the
uncertainty principle and it incorporates gravitational implications
in general relativity (GR). This model considers changes in
spacetime and the content of matter to explain DE as determined by
the universe. Cai \cite{4} first introduced the original ADE model
to study the rapid expansion of the cosmos. The expression for
energy density, $\rho_{D}=3n^{2}M^{2}_{p}T^{-2}$, includes the age
($T$) of the cosmos, $M^{2}_{p}$ is the Planck mass and the
numerical value $3n^{2}$ is used to accommodate for some
uncertainties. However, this framework has certain limitations that
cannot be explained by the matter-dominated era of the universe. Wei
and Cai \cite{5} proposed a novel framework in the form of the new
ADE (NADE) model, which replaces the age of the universe with
conformal time. The coincidence problem is naturally solved by this
model \cite{6}.

Recent interest in cosmology has focused on the reconstruction
scenario involving different DE models. Setare \cite{10} explored
the NADE model in $f(\mathcal{R})$ gravity ($\mathcal{R}$ is the
Ricci scalar) and found evidence about the possible existence of the
universe with phantom-like characteristics. Jamil and Saridakis
\cite{11} proposed the NADE model in the context of Horava-Lifshitz
gravity, demonstrating its consistency with observations regarding
the rapid expansion of the cosmos. Li et al. \cite{12} investigated
the behavior of the NADE as a rolling tachyon to examine its both
potential and dynamics as a scalar field. Zhang et al. \cite{13}
studied the cosmic evolution of the NADE model with interaction
between DE and matter component through statefinder parameter.
Houndjo and Piattella \cite{15} analyzed the numerical
reconstruction of the $f(\mathcal{R},\mathcal{T})$ gravity
($\mathcal{T}$ represents the trace of the EMT) that shows the
features of HDE models. They examined the HDE and NADE models and
constructed the corresponding $f(\mathcal{R},\mathcal{T})$ gravity
model as an alternative representation without the need for
additional DE components.

Sharif and Jawad \cite{16} investigated the mysterious
characteristics of HDE and NADE models in the framework of GR. Fayaz
et al. \cite{17} used a Bianchi type-I cosmological model in the
framework of reconstructed $f(\mathcal{R},\mathcal{T})$ gravity to
investigate the phantom and quintessence phases of cosmic evolution
in HDE and NADE models. Setare et al. \cite{18} computed the
perturbed quantities for the NADE model and evaluated the results of
the standard cold DM (CDM) model. Sharif and Saba \cite{19} examined
the cosmic dynamics of the reconstructed models using the phase
planes and the cosmic diagnostic parameters. Pourbagher and Amani
\cite{20} analyzed the cosmological parameters and found that the
total entropy variation increases as time progresses under
thermodynamic equilibrium for specific free parameters in NADE model
with $f(\mathcal{T},\mathcal{B})$ theory, where $\mathcal{B}$ is
boundary term.

The concept of GR is based on Riemannian geometry and asserts that
the affine connection on the spacetime manifold must align with the
metric, known as the Levi-Civita connection \cite{3a}. However,
there can exist multiple options for an affine connection on any
manifold. It is theoretically viable to explore gravitational
theories using non-Riemannian geometry in which the curvature,
torsion, and non-metricity all have non-zero values. When choosing a
connection for which both curvature as well as non-metricity
disappear, but allowing for some variation in torsion, it becomes
feasible to formulate the teleparallel equivalent of GR \cite{21}.
Considering a flat spacetime manifold without torsion but with a
non-zero nonmetricity, the symmetric teleparallel formulation of GR
is obtained \cite{22}. The $f$-theories are a category of modified
theories and $f(\mathcal{R})$ gravity is focused on the Ricci scalar
of the Levi-Civita connection. The $f(\mathbb{T})$ \cite{23} and
$f(\mathcal{Q})$ \cite{24} theories of gravity ($\mathbb{T}$ and
$\mathcal{Q}$ represent the torsion scalar and non-metricity,
respectively) address the curvature-less Weitzenb$\ddot{o}$ck
connection. The $f(\mathcal{R})$, $f(\mathbb{T})$, and
$f(\mathcal{Q})$ theories represent entirely different gravitational
frameworks each typically offering a unique gravitational evolution.
All three theories have shared the features in which each enables a
mini-superspace depiction in the study of cosmology. For a
non-linear function, the theory of gravity described by
$f(\mathcal{R})$ is the fourth-order, while the $f(\mathbb{T})$ and
$f(\mathcal{Q})$ theories are of the second-order.

Consequently, the existence of a scalar field resulting from the
higher-order derivatives ($f(\mathcal{R})$ gravity) raised the
degree of freedom, which results in the theory being equal to a
scalar-tensor theory. We analyze the $f(\mathcal{Q})$ theory, an
extension of the symmetric teleparallel GR (STGR) where gravity
arises from the non-metricity. The theory is motivated by the need
to explore its various underlying factors including theoretical
consequences, consistency with observed data and its significance in
cosmic contexts. This theory investigates theoretical effects based
on cosmic domains and observational evidence. The metric tensor in
$f(\mathcal{Q})$ theory has a non-zero covariant derivative which
can be described using a new geometric variable called
non-metricity. In non-Riemannian gravity, the field strengths
include the non-metricity tensor $\mathcal{Q}_{\zeta\xi}$, torsion
scalar $\mathbb{T}$ and curvature tensor $\mathcal{R}_{\zeta\xi}$.
The classification of spacetimes and related theories are discussed
in Table \textbf{1}.
\begin{table}\caption{Classification of spacetimes}
\begin{center}
\begin{tabular}{|c|c|c|}
\hline Relations  & Spacetimes & physical representations
\\
\hline $\mathcal{Q}_{\zeta\xi}=0,~ \mathbb{T}=0,~
\mathcal{R}_{\zeta\xi}=0$ & Minkowski & Special Relativity
\\
$\mathcal{Q}_{\zeta\xi}=0,~ \mathbb{T}=0,~
\mathcal{R}_{\zeta\xi}\neq0$ & Riemannian & General Relativity
\\
$\mathcal{Q}_{\zeta\xi}=0,~ \mathbb{T}\neq0,~
\mathcal{R}_{\zeta\xi}=0$ & Weitzenbock & Teleparallel Gravity
\\
$\mathcal{Q}_{\zeta\xi}\neq0,~ \mathbb{T}=0,~
\mathcal{R}_{\zeta\xi}=0$ & & Symmetric Teleparallel
\\
$\mathcal{Q}_{\zeta\xi}\neq0,~ \mathbb{T}=0,~
\mathcal{R}_{\zeta\xi}\neq0$ & Riemann-Weyl & Einstein-Weyl
\\
$\mathcal{Q}_{\zeta\xi}=0,~ \mathbb{T}\neq0,~
\mathcal{R}_{\zeta\xi}\neq0$ & Riemann-Cartan & Einstein-Cartan
\\
$\mathcal{Q}_{\zeta\xi}\neq0,~ \mathbb{T}\neq0,~
\mathcal{R}_{\zeta\xi}\neq0$ & Non-Riemannian & Einstein-Cartan-Weyl
\\
\hline
\end{tabular}
\end{center}
\end{table}

Recent studies on $f(\mathcal{Q})$ gravity have uncovered cosmic
challenges and observational limitations that can be used to
demonstrate variations from the standard CDM model. Lu et al.
\cite{25} researched the cosmic properties in STGR and described
that the universe's geometric nature contributes to its accelerating
expansion. Lazkoz et al. \cite{26} studied the cosmic evolution
using $f(\mathcal{Q})$ as polynomial functions of the redshift.
Frusciante \cite{31} proposed a particular model in this gravity.
This model shared similarities with the $\Lambda$CDM model at a
fundamental level.

Mandal and Sahoo \cite{33} investigated the Hubble, Pantheon sample
and the equation of state (EoS) parameters. The results of the
standard CDM model are different from the $f(\mathcal{Q})$ model,
which suggest quintessential behavior. Myrzakulov et al. \cite{34}
conducted a study on the cosmography of ghost DE and pilgrim DE in
this theory. A recent investigation explored methods for
parameterizing the effective EoS parameter in this context. Lymperis
\cite{35} analyzed the same theoretical framework to investigate the
cosmological implications of the effective DE sector. Solanki et al.
\cite{36} found that the source of DE could be explained by the
geometric expansion of GR. Koussour et al. \cite{37} examined the
properties of cosmic parameters in this gravity. In recent papers
\cite{38}, we have developed generalized ghost DE and generalized
ghost pilgrim DE models in the same gravity using the correspondence
principle in a non-interacting framework. Additionally, we have
examined the pilgrim and generalized ghost pilgrim DE models for the
non-interacting scenario \cite{39}. These models effectively
replicate various cosmic epochs and align well with the latest
observational data.

This paper uses the correspondence scheme to reconstruct the
interacting case of the NADE $f(\mathcal{Q})$ model. Investigating
the evolution of the universe involves studying the EoS parameter as
well as analyzing the squared speed of sound and phase planes. The
article is structured as follows. In section \textbf{2}, we give a
summary of $f(\mathcal{Q})$ gravity and its significance for
cosmological studies. In section \textbf{3}, the impacts of combined
DE and CDM interaction are examined about the red-shift parameter.
Furthermore, a method is employed to establish a link between NADE
and $f(\mathcal{Q})$ gravity to devise a NADE $f(\mathcal{Q})$
model. The purpose of section \textbf{4} is to examine this model's
evolution using cosmographic analysis. Our results are summarized in
section \textbf{5}.

\section{A Brief Overview of $f(\mathcal{Q})$ Gravity}

In this section, assuming the properties of the affine connection
essentially define a metric-affine geometry \cite{40}. The
gravitational potential can be considered as a value extended by the
metric tensor $g_{\zeta\xi}$. In this particular context, a
fundamental theorem in differential geometry asserts that the
overall affine connection can be broken down into three distinct and
separate components \cite{41}
\begin{equation}\label{1}
\hat{\Gamma}^{\lambda}_{\zeta\xi}={\Gamma}^{\lambda}_{\zeta\xi}
+\mathcal{C}^{\lambda}_{\;\zeta\xi}+\mathcal{L}^{\lambda}_{\;\zeta\xi},
\end{equation}
where $\Gamma^{\lambda}_{\zeta\xi}=\frac{1}{2}g^{\lambda\sigma}
(g_{\sigma\xi,\zeta}+g_{\sigma\zeta,\xi}-g_{\zeta\xi,\sigma})$
represents the Levi-Civita connection. The term
$\mathcal{C}^{\lambda}_{\;\zeta\xi}=\hat{\Gamma}^{\lambda}_{[\zeta\xi]}
+g^{\lambda\sigma}g_{\zeta\kappa}\hat{\Gamma}^{\kappa}_{[\xi\sigma]}
+g^{\lambda\sigma}g_{\xi\kappa}\hat{\Gamma}^{\kappa}_{[\zeta\sigma]}$
denotes the contortion, characterized by the torsion tensor
$\mathcal{T}_{\zeta\xi}^{\alpha}=2\hat{\Gamma}^{\alpha}_{[\zeta\xi]}$,
and lastly, the disformation $\mathcal{L}^{\lambda}_{\;\zeta\xi}$ is
determined by
\begin{equation}\label{2}
\mathcal{L}^{\lambda}_{\;\zeta\xi}=\frac{1}{2}g^{\lambda\sigma}(\mathcal{Q}_{\xi\zeta\sigma}
+\mathcal{Q}_{\zeta\xi\sigma}-\mathcal{Q}_{\lambda\zeta\xi}),
\end{equation}
which is expressed in relation to the non-metricity tensor
$\mathcal{Q}_{\xi\zeta\sigma}=\nabla_{\sigma}g_{\zeta\xi}\neq 0$.
This study will concentrate on a non-metric geometry which is
characterized solely by its non-metricity tensor
$\mathcal{Q}_{\xi\zeta\sigma}$, without any torsion or curvature.
This innovative method has undergone many cosmological experiments
and its investigation provided valuable understanding of the
universe's late accelerated expansion. In the framework of different
modified gravity theories, we start by considering the concept of
extending $\mathcal{Q}$-gravity in a similar way as $f(\mathcal{R})$
theory has been generalized.

Considering the integral action of $f(\mathcal{Q})$ gravity as
\cite{22}
\begin{equation}\label{3}
S=\int\left(\frac{1}{2k}f(\mathcal{Q})+L_{m}\right) \sqrt{-g}d^{4}x,
\end{equation}
while the matter lagrangian density is denoted by $L_{m}$, $g$
represents the determinant of the metric tensor and $f(\mathcal{Q})$
represents an arbitrary function of $\mathcal{Q}$, which can be
described as
\begin{equation}\label{4}
\mathcal{Q}=-g^{\zeta\xi}(\mathcal{L}^{\mu}_{~\nu\zeta}\mathcal{L}^{\nu}_{~\xi\mu}
-\mathcal{L}^{\mu}_{~\nu\mu}\mathcal{L}^{\nu}_{~\zeta\xi}).
\end{equation}
Since the Levi-Civita connection in symmetric connections can be
expressed in terms of the disfomation tensor as
$\Gamma^{\lambda}_{\zeta\xi}=-\mathcal{L}^{\lambda}_{\;\zeta\xi}$,
thus we have
\begin{equation}\label{5}
\mathcal{L}^{\lambda}_{\;\zeta\xi}=-\frac{1}{2}g^{\lambda\sigma}
(\nabla_{\zeta}g_{\sigma\xi}+\nabla_{\xi}g_{\sigma\zeta}
-\nabla_{\sigma}g_{\zeta\xi}).
\end{equation}
The superpotential can be defined as a function of $\mathcal{Q}$
given by
\begin{equation}\label{7}
\mathcal{P}^{\mu}_{\;\zeta\xi}=-\frac{1}{2}\mathcal{L}^{\mu}_{\;\zeta\xi}
+\frac{1}{4}(\mathcal{Q}^{\mu}-\tilde{\mathcal{Q}}^{\mu})g_{\zeta\xi}-
\frac{1}{4} \delta ^{\mu}\;_{({\zeta}}\mathcal{Q}_{\xi)}.
\end{equation}
A different type of superpotential is described using Eq.\eqref{2}
in \eqref{7} as
\begin{equation}\nonumber
\mathcal{P}^{\mu\zeta\xi}=\frac{1}{4}\big[-\mathcal{Q}^{\mu\zeta\xi}+
\mathcal{Q}^{\zeta\mu\xi}
+\mathcal{Q}^{\xi\mu\zeta}+\mathcal{Q}^{\zeta\mu\xi}-\tilde{Q}_{\mu}g^{\zeta\xi}
+\mathcal{Q}^{\mu}g^{\zeta\xi}
-\frac{1}{2}(\mathcal{Q}^{\xi}g^{\mu\zeta}+\mathcal{Q}^{\zeta}g^{\mu\xi})\big],
\end{equation}
\begin{equation}\label{8}
\mathcal{Q}=-\mathcal{Q}_{\mu\zeta\xi}\mathcal{P}^{\mu\zeta\xi}=
-\frac{1}{4}(-\mathcal{Q}^{\mu\xi\rho}\mathcal{Q}_{\mu\xi\rho}
+2\mathcal{Q}^{\mu\xi\rho}\mathcal{Q}_{\rho\mu\xi}-2\mathcal{Q}^{\rho}
\tilde{\mathcal{Q}}_{\rho}+\mathcal{Q}^{\rho}\mathcal{Q}_{\rho}),
\end{equation}
where
\begin{equation}\label{6}
\mathcal{Q}_{\mu}=\mathcal{Q}^{~\zeta}_{\mu~\zeta},\quad
\tilde{\mathcal{Q}}_{\mu}=\mathcal{Q}^{\zeta}_{~\mu\zeta}.
\end{equation}
Choosing $k=1$ for simplicity gives the field equations of
$f(\mathcal{Q})$ gravity, given as follows
\begin{equation}\label{9}
\frac{-2}{\sqrt{-g}}\nabla_{\zeta}(f_{\mathcal{Q}}\sqrt{-g}
P^{\mu}_{~\zeta\xi})-\frac{1}{2}f g_{\zeta\xi}-f_{\mathcal{Q}}
(P_{\zeta\mu\nu}\mathcal{Q}_{\xi}^{~\mu\nu}-2\mathcal{Q}^{\mu\nu}_{~~~\zeta}
P_{\mu\nu\xi})= \mathcal{T}_{\zeta\xi},
\end{equation}
where the EMT for matter is expressed by $\mathcal{T}_{\zeta\xi}$
and $f_{\mathcal{Q}}=\frac{\partial f(\mathcal{Q})}{\partial
\mathcal{Q}}$.

\section{Restructuring the NADE $f(\mathcal{Q})$ Model}

In this section, we reconstruct the NADE $f(\mathcal{Q})$ gravity
model through correspondence principle by using flat
Friedmann-Robertson-Walker (FRW) universe model given as
\begin{equation}\label{10}
ds^{2}=-dt^{2}+a^{2}(t)(dx^{2}+dy^{2}+dz^{2}),
\end{equation}
where the scale factor is represented by $a(t)$. The EMT for a
perfect fluid is defined as
$\bar{\mathcal{T}}_{\;\zeta\xi}=(\rho_{m}+p_{m})u_{\zeta}u_{\xi}+p_{m}
g_{\zeta\xi}$, with $\rho_{m}$ and $p_{m}$ representing the
thermodynamic energy density and isotropic pressure, respectively,
$u_{\zeta}$ represents the the four-velocity field. We derive the
Friedmann equations in $f(\mathcal{Q})$ gravity as
\begin{equation}\label{11}
3H^{2}=\rho_{m}+\rho_D,\quad 2\dot{H}+3H^{2}=p_{m}+p_D,
\end{equation}
where the derivative with respect to $t$ is indicated by an upper
dot in the Hubble function $H=\frac{\dot{a}}{a}$. The density and
pressure of the DE are provided as
\begin{eqnarray}\label{12}
\rho_D&=&\frac{f}{2}-6H^{2}f_{\mathcal{Q}},\\\label{12a}
p_D&=&\frac{f}{2}+2f_{\mathcal{Q}}\dot{H}+2Hf_{\mathcal{Q}
\mathcal{Q}}+6H^{2}f_{\mathcal{Q}},
\end{eqnarray}
here $\Omega_{D}$ and $\Omega_{m}$ are the two fractional energy
densities expressed as follows
\begin{equation}\label{13}
\Omega_{D}=\frac{\rho_{D}}{\rho_{cr}}=\frac{\rho_{D}}{3H^{2}}, \quad
\Omega_{m}=\frac{\rho_{m}}{\rho_{cr}}=\frac{\rho_{m}}{3H^{2}},
\end{equation}
one can represent $1$ as the sum of $\Omega_{D}$ and $\Omega_{m}$,
where $\rho_{cr}$ denotes the critical density.

Suppose the interplay between two fluid components, namely the DE
and DM. As a result, when considering both fluids together, their
respective energy densities do not individually remain constant but
instead assume a specific form in the interacting scenario
\begin{equation}\label{14}
\dot{\rho}_{m}+3H(\rho_{m}+p_{m})=\Gamma,\quad
\dot{\rho}_{D}+3H(\rho_{D}+p_{D})=-\Gamma,
\end{equation}
the interaction term in this case is denoted by $\Gamma$. It is
clear that for energy transfer from DE to DM to occur, $\Gamma$ must
be positive. The value of $\Gamma$ is simply determined as the
product of $H$ and $\rho_{D}$, since it is the inverse of time
evolution. Here we take $\Gamma=3\psi H(\rho_{m}+p_{D})=3\psi
H\rho_{D}(1+\chi)$ \cite{34}, where the coupling constant $\psi$
indicates the strength of the interaction between DE and DM. By
carefully examining the role of $\psi$, we have found that varying
its value significantly influences the universe expansion rate,
highlighting its critical role in cosmological evolution. Our
results demonstrate how the interaction between these components
affects the dynamics of the universe, emphasizing the importance of
this factor in the broader analysis of cosmic evolution. The
parameter $\chi$ is defined as
\begin{equation}\label{15}
\chi=\frac{\rho_{m}}{\rho_{D}}=\frac{\Omega_{m}}{\Omega_{D}}
=\frac{1-\Omega_{D}}{\Omega_{D}}.
\end{equation}
We can represent $\omega_{D}$ using the parameters that have been
established previously \cite{38}
\begin{equation}\label{16}
\omega_{D}=-\frac{1}{2-\Omega_{D}}\bigg(1+\frac{2\psi}{\Omega_{D}}\bigg).
\end{equation}
Substituting the age of the universe $T$ with the conformal time
$\eta$ in the energy density of the ADE model, we obtain the energy
density of the NADE model
\begin{equation}\nonumber
\rho_D= \frac{3n^{2}M^{2}_{p}}{\eta ^{2}},\quad
\eta=\int\frac{dt}{a(t)},
\end{equation}
where $n$ is an arbitrary constant.

This model offers an alternative explanation to the accelerated
expansion of the cosmos using the age of the universe as a measure
of cosmic energy density. For simplification of subsequent
calculations, we set $M^{2}_{p}=1$ and impose the restriction $n>1$
to obtain
\begin{equation}\label{17}
\rho_D= \frac{3n^{2}}{\eta ^{2}}.
\end{equation}
Taking the equivalent densities equal to each other, we demonstrate
the connection between NADE and the $f(\mathcal{Q})$ gravity
\cite{43}. From Eqs.\eqref{12} and \eqref{17}, it is clear that
\begin{equation}\label{18}
\frac{f}{2}-6H^{2}f_{\mathcal{Q}}=\frac{3 n^{2}}{\eta ^{2}}.
\end{equation}
This is the first-order linear differential equation in
$\mathcal{Q}$ and its solution is
\begin{equation}\label{19}
f(\mathcal{Q})=c \sqrt{\mathcal{Q}}+\frac{12 n^{2}}{\eta ^{2}},
\end{equation}
where $c$ represents the integration constant.

Now, we express this solution \eqref{19} in relation to the redshift
parameter $z$. We represent the scale factor using a power-law
formulation expressed as $a(t)=a_{0}t^{j}$, where $j$ and $a_0$ are
arbitrary constants, with the current value of $a_0$ being equal to
1. The deceleration parameter is characterized by
$q=-\frac{a\ddot{a}}{\dot{a}^{2}}=-1+\frac{1}{j}$. Replacing the
value of $j$ in the function $a(t)$, we have
\begin{equation}\label{20}
a(t)=t^{\frac{1}{1+q}},
\end{equation}
where $q=-0.832^{+0.091}_{-0.091}$ \cite{44}, with $q>-1$ indicating
that the universe is expanding. This value reflects the acceleration
of the universe at the present time. Utilizing this scale factor, we
can express
\begin{equation}\label{21}
H=(1+q)^{-1}t^{-1},\quad H_{0}=(1+q)^{-1}t_{0}^{-1}.
\end{equation}
This suggests that $q$ and $H_{0}$ are the parameters that determine
the expansion of the universe. When we evaluate the connection
between $z$ and the scale factor, we obtain
\begin{equation}\label{22}
H=H_{0}\Psi^{1+q} , \quad \dot{H} =-H_{0}\Psi^{2+2q},
\end{equation}
where $\Psi=1+z$. The value of $\mathcal{Q}$ is calculated by
\cite{38}
\begin{equation}\nonumber
\mathcal{Q}=6H^{2}.
\end{equation}
Applying the value of $H$, we obtain
\begin{equation}\label{23}
\mathcal{Q}=6H_{0}^{2}\Psi^{2+2q}.
\end{equation}
When we substitute this value in Eq.\eqref{19}, We can express the
solution in terms of $z$ as follows
\begin{equation}\label{24}
f(\mathcal{Q})=\sqrt{6} c \sqrt{H_{0}^2 \Psi^{2q+2}}+\frac{12 n^2
q^2 \Psi^{2 q}}{(q+1)^2}.
\end{equation}

For the purpose of analysis, we use three fixed values of $n=11,
11.4$ and $11.8$ to explore the graphical behavior in the
$f(\mathcal{Q})$ theory. If we change the value of $n$, it has a
distinct impact on these graphical representations. These values
were chosen to provide a close examination of the model's behavior
under slight variations, allowing us to analyze the stability and
consistency of the results. The behavior of the graphs with these
values is favorable, as it leads to good representations in
parametric graphs (phase-planes). We have considered the current
value of the Hubble constant $H_{0}$ as $70 Km s^{-1} Mpc^{-1}$,
which is widely accepted based on recent observational data. This
value is used throughout the analysis to ensure consistency in the
calculated quantities. Any variation in the Hubble constsnt would
influence the results, but our choice reflects the present-day
accepted value from cosmological observations. Additionally, We
arbitrarily set the constant of integration $c=2$, which negligibly
impacts the graphical behavior of the plots.
\begin{figure}
\epsfig{file=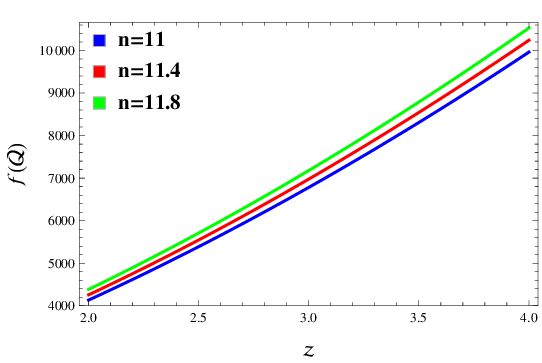,width=0.52\linewidth}
\epsfig{file=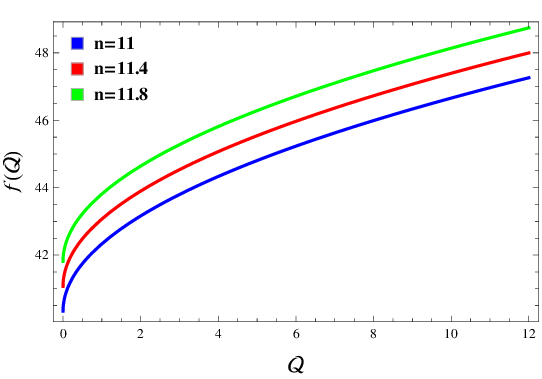,width=0.5\linewidth}\caption{Graph of
$f(\mathcal{Q})$ against $z$ and $\mathcal{Q}$.}
\end{figure}

Figure \textbf{1} demonstrates that the reconstructed NADE model
consistently stays positive and rises with both $z$ and $Q$ for all
chosen values of $n$. We also examine the characteristics of
$\rho_D$ and $p_D$ in the context of NADE reconstructed
$f(\mathcal{Q})$ gravity model. Applying Eq.\eqref{19} to \eqref{12}
and \eqref{12a}, we derive
\begin{eqnarray}\nonumber
\rho_D&=&\frac{6 n^2}{\eta ^2}-\frac{1}{2} c \bigg(\sqrt{6}
H-\sqrt{\mathcal{Q}}\bigg),\\\nonumber p_D&=&\frac{c \eta ^2
\bigg(\mathcal{Q} (2 \dot{H}-\mathcal{Q})+6 H^2
\mathcal{Q}-H\bigg)-12 n^2 \mathcal{Q}^{3/2}}{2 \eta ^2
\mathcal{Q}^{3/2}},
\end{eqnarray}
where $\sigma=q+1$ for further simplification. In terms of redshift
parameter, these equations take the following form
\begin{eqnarray}\label{25}
\rho_D&=&\sqrt{\frac{3}{2}} c \bigg(\sqrt{H_{0}^2 \Psi^{2
q+2}}-H_{0} \Psi^\sigma\bigg)+\frac{6 n^2 q^2 \Psi^{2 q}}{\sigma^2},
\\\nonumber
p_D &=&\bigg[q^2 \Psi^{2 q} \bigg\{-\bigg[\bigg(c H_{0} \sigma^2
\Psi^{1-q} \big(12 H_{0}^2 \Psi^{3
q+3}+1\big)\bigg)\big(q^2\big)^{-1}\bigg]
\\
\label{26} &-&72 \sqrt{6} n^2\bigg(H_{0}^2 \Psi^{2
q+2}\bigg)^{3/2}\bigg\}\bigg] \bigg[12 \sqrt{6} \sigma^2
\bigg(H_{0}^2 \Psi^{2 q+2}\bigg)^{3/2}\bigg]^{-1}.
\end{eqnarray}
Figure \textbf{2} shows how the reconstructed NADE $f(\mathcal{Q})$
gravity behaves with $z$. For all values of $n$, the reconstructed
NADE $f(\mathcal{Q})$ gravity has an exponentially increasing
$\rho_D$. The quantity $p_D$ indicates a decreasing pattern and
continuously shows negative behavior, which corresponds with the DE
behavior.
\begin{figure}
\epsfig{file=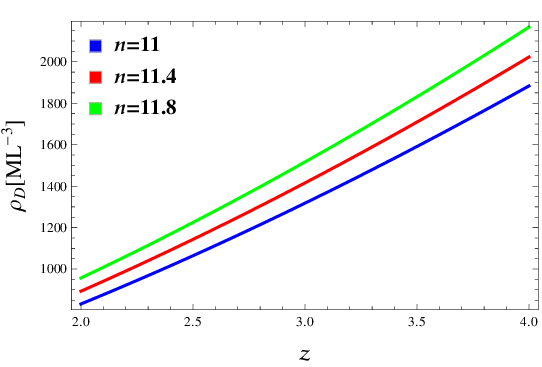,width=0.51\linewidth}
\epsfig{file=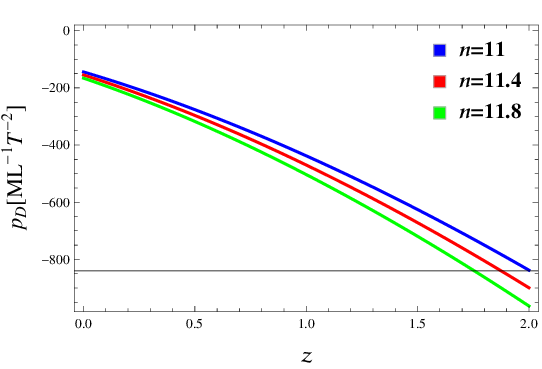,width=0.51\linewidth}\caption{Graphs of $\rho_D$
and $p_D$ against $z$.}
\end{figure}

\section{Cosmographic Analysis}

In this section, we perform cosmographic analysis on the EoS
parameter and phase planes for the reconstructed NADE
$f(\mathcal{Q})$ gravity model in an interacting scenario to
investigate the universe evolution. We also explore $\nu_{s}^{2}$ to
analyze the stability of this model.

In this context, the negative values of the coupling constant were
chosen because they provided consistent and meaningful results for
the model we are exploring. While positive values can lead to
changes in graphical behavior, they may not achieve the same level
of consistency with observational data. As noted by Feng et al.
\cite{1abc}, a small coupling constant is necessary to align with
observations and addresses the coincidence problem. Our analysis
shows that employing a small coupling constant, even if negative,
helps avoid this problem while remaining compatible with current
observations.

\subsection{Equation of State Parameter}

The equation of state parameter
($\omega_{D}=\frac{p_{D}}{\rho_{D}}$) for DE is essential in
characterizing the cosmic inflation phase and the subsequent
expansion of the cosmos. We study the condition for the universe
undergoing acceleration, which happens when the EoS
$\omega_{D}<-\frac{1}{3}$. When $\omega_{D}=-1$, it represents the
cosmological constant. However, the cases $\omega_{D}=\frac{1}{3}$
and $\omega_{D}=0$ denote radiation-dominated and matter-dominated
eras, respectively. Furthermore, the phantom situation arises when
$\omega_{D}<-1$, while $-1<\omega_{D}<-\frac{1}{3}$ leads to
quintessence phase of the universe expansion. Referring to
Eq.\eqref{16}, we can derive
\begin{eqnarray}\nonumber
\omega_{D}&=&-\bigg\{\eta ^2 \mathcal{Q} \bigg(\eta ^2
\bigg(\sqrt{6} c H-c \sqrt{\mathcal{Q}}+2 \mathcal{Q} \psi \bigg)-12
n^2\bigg)\bigg\}\bigg\{\bigg(
\bigg(\sqrt{\mathcal{Q}}-\sqrt{6} H\bigg)\\
\label{27}&\times&c \eta ^2+12 n^2\bigg) \bigg(\eta ^2 \bigg(c
\big(\sqrt{\mathcal{Q}}-\sqrt{6} H\big)-2 \mathcal{Q}\bigg)+12
n^2\bigg)\bigg\}^{-1},
\end{eqnarray}
while in the context of $z$, this is expressed as
\begin{eqnarray}\nonumber
\omega_{D}&=& -\bigg[6 H_{0}^2 \sigma^2 \Psi^2
\bigg\{\bigg[\bigg(\sigma^2 \bigg(12 H_{0}^2 \psi  \Psi^{2
q+2}-\sqrt{6} c \bigg(\sqrt{H_{0}^2
\Psi^{2 q+2}}-H_{0} \Psi^{q+1}\bigg)\bigg)\\
\nonumber&\times& \Psi^{-2 q}\bigg)\big(q^2\big)^{-1}\bigg]-12
n^2\bigg\}\bigg]\bigg[q^2 \bigg\{\bigg[\bigg(\sqrt{6}\Psi^{-2
q} \bigg(\sqrt{H_{0}^2 \Psi^{2 q+2}}-H_{0} \Psi^{\sigma}\bigg)\\
\nonumber&\times&c \sigma^2\bigg)\big(q^2\big)^{-1}\bigg]+12
n^2\bigg\}\bigg\{12 n^2-\bigg[\bigg( \bigg(12 H_{0}^2 \Psi^{2
q+2}-\sqrt{6} c \bigg(\sqrt{H_{0}^2 \Psi^{2 q+2}}\\
\nonumber&-&H_{0} \Psi^{\sigma}\bigg)\bigg)\sigma^2 \Psi^{-2
q}\bigg)(q^2)^{-1}\bigg]\bigg\}\bigg]^{-1}.
\end{eqnarray}
Figure \textbf{3} demonstrates the dynamical evolution of EoS in the
NADE $f(\mathcal{Q})$ gravity for various values of $n$ and $\psi$.
It exhibits values greater than $-1$ and less than $-\frac{1}{3}$,
specifically expressed as $-1<\omega_{D}<-\frac{1}{3}$. This
suggests the presence of quintessence field DE in this model.
\begin{figure}
\epsfig{file=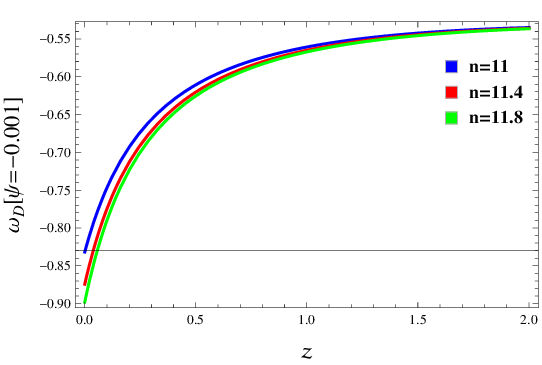,width=.5\linewidth}
\epsfig{file=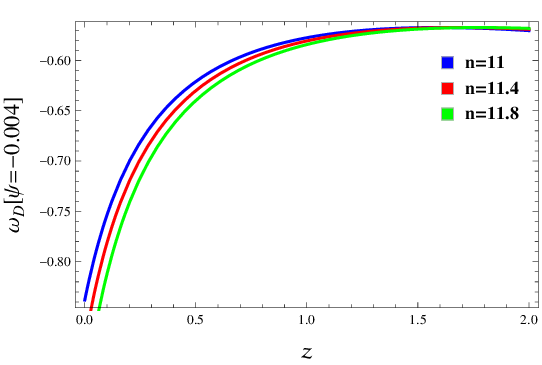,width=.5\linewidth}\center \epsfig
{file=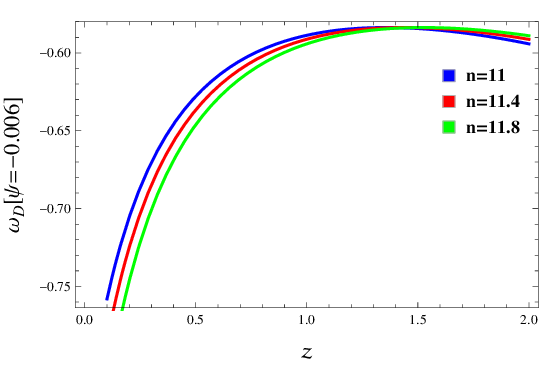,width=.5\linewidth} \caption{Plots of $\omega_{D}$
versus $z$.}
\end{figure}

\subsection{The ($\omega_{D}-\omega^{\prime}_{D}$)-Plane}

Here, we make use of the phase plane
$(\omega_{D}-\omega^{\prime}_{D})$, where $\omega^{\prime}_{D}$
represents the evolutionary behavior of $\omega_{D}$ and prime
indicates the derivative with respect to $Q$. Caldwell and Linder
\cite{45} introduced this cosmic framework to explore the
quintessence DE paradigm, which can be divided into freezing
$(\omega_{D}<0,~\omega^{\prime}_{D}<0 )$ and thawing
$(\omega_{D}<0,~\omega^{\prime}_{D}>0)$ scenarios. The current
cosmic expansion model is represented by the freezing region, which
indicates a more rapid phase in comparison to thawing region.
Differentiating Eq.\eqref{27} with respect to $\mathcal{Q}$ gives us
\begin{eqnarray}\nonumber
\omega^{\prime}_{D}&=&-\frac{\eta ^2 \bigg(\eta ^2 \big(\sqrt{6} c
H-c \sqrt{\mathcal{Q}}+2 \mathcal{Q} \psi \big)-12
n^2\bigg)}{\bigg(c \eta ^2 \big(\sqrt{\mathcal{Q}}-\sqrt{6}
H\big)+12 n^2\bigg) \bigg(\eta ^2 \big(c
\big(\sqrt{\mathcal{Q}}-\sqrt{6} H\big)-2 \mathcal{Q}\big)+12 n^2\bigg)}\\
\nonumber&+&\frac{c \eta ^4 \sqrt{\mathcal{Q}} \bigg(\eta ^2
\big(\sqrt{6} c H-c \sqrt{\mathcal{Q}}+2 \mathcal{Q} \psi \big)-12
n^2\bigg)}{2 \bigg(c \eta ^2 \big(\sqrt{\mathcal{Q}}-\sqrt{6}
H\big)+12 n^2\bigg)^2 \bigg(\eta ^2 \big(c
\big(\sqrt{\mathcal{Q}}-\sqrt{6} H\big)-2 \mathcal{Q}\big)+12
n^2\bigg)}\\
\nonumber&+&\frac{\eta ^4 \mathcal{Q} \bigg(\frac{c}{2
\sqrt{\mathcal{Q}}}-2\bigg) \bigg(\eta ^2 \big(\sqrt{6} c H-c
\sqrt{\mathcal{Q}}+2 \mathcal{Q} \psi \big)-12 n^2\bigg)}{\bigg(c
\eta ^2 \big(\sqrt{\mathcal{Q}}-\sqrt{6} H\big)+12 n^2\bigg)
\bigg(\eta ^2 \big(c \big(\sqrt{\mathcal{Q}}-\sqrt{6} H\big)-2
\mathcal{Q}\big)+12 n^2\bigg)^2}\\
\nonumber&-&\frac{\eta ^4 \mathcal{Q} \bigg(2 \psi -\frac{c}{2
\sqrt{\mathcal{Q}}}\bigg)}{\bigg(c \eta ^2
\bigg(\sqrt{\mathcal{Q}}-\sqrt{6} H\bigg)+12 n^2\bigg) \bigg(\eta ^2
\bigg(c \bigg(\sqrt{\mathcal{Q}}-\sqrt{6} H\bigg)-2
\mathcal{Q}\bigg)+12 n^2\bigg)}.
\end{eqnarray}
In terms of $z$, we can write as follows
\begin{eqnarray}\nonumber
\omega^{\prime}_{D}&=& \bigg[\bigg\{\sigma^2 \Psi^{-4 q} \bigg(-2
q^2 \bigg(\bigg[\bigg\{\sqrt{6} c \sigma^2 \bigg(\sqrt{H_{0}^2
\Psi^{2 q+2}}-H_{0} \Psi^{\sigma}\bigg) \Psi^{-2
q}\bigg\}\big\{q^2\big\}^{-1}\bigg]\\
\nonumber&+&12 n^2\bigg) \bigg(12 n^2-\bigg[\bigg\{\sigma^2 \bigg(12
H_{0}^2 \Psi^{2 q+2}-\sqrt{6} c \bigg(\sqrt{H_{0}^2 \Psi^{2
q+2}}-H_{0} \Psi^{\sigma}\bigg)\bigg)\\
\nonumber&\times&\Psi^{-2 q}\bigg\}\big\{q^2\big\}^{-1}\bigg]\bigg)
\bigg(\bigg[\bigg\{\bigg(12 H_{0}^2 \Psi^{2 q+2} \psi -\sqrt{6} c
\bigg(\sqrt{H_{0}^2 \Psi^{2 q+2}}-H_{0}
\Psi^{\sigma}\bigg)\bigg)\\
\nonumber&\times&\sigma^2\Psi^{-2
q}\bigg\}\big\{q^2\big\}^{-1}\bigg]-12 n^2\bigg) \Psi^{2 q}+12
H_{0}^2 \sigma^2 \bigg(\bigg[\bigg\{\sqrt{6} c \sigma^2
\bigg(\sqrt{H_{0}^2 \Psi^{2 q+2}}\\
\nonumber&-&H_{0} \Psi^{\sigma}\bigg) \Psi^{-2
q}\bigg\}\big\{q^2\big\}^{-1}\bigg]+12 n^2\bigg)
\bigg(\bigg[\bigg\{\sigma^2 \Psi^{-2 q} \bigg(12 H_{0}^2 \Psi^{2
q+2} \psi -\sqrt{6} c \\
\nonumber&\times&\bigg(\sqrt{H_{0}^2 \Psi^{2 q+2}}-H_{0}
\Psi^{\sigma}\bigg)\bigg)\bigg\}\big\{q^2\big\}^{-1}\bigg]-12
n^2\bigg) \bigg(\frac{c}{2 \sqrt{6} \sqrt{H_{0}^2 \Psi^{2
q+2}}}-2\bigg)
\\
\nonumber&\times& \Psi^{2 q+2}\bigg(\bigg[\bigg\{
\bigg(\sqrt{H_{0}^2\Psi^{2 q+2}} -H_{0} \Psi^{\sigma}\bigg)\sqrt{6}
c \sigma^2 \Psi^{-2 q}\bigg\}\big\{q^2\big\}^{-1}\bigg]+12
n^2\bigg) \\
\nonumber&-&12\sigma^2 H_{0}^2 \bigg(12 n^2-\bigg[\bigg\{\bigg(12
H_{0}^2 \Psi^{2 q+2}-\sqrt{6} c \bigg(\sqrt{H_{0}^2 \Psi^{2
q+2}}-H_{0}
\Psi^{\sigma}\bigg)\bigg)\\
\nonumber&\times&\Psi^{-2
q}\sigma^2\bigg\}\big\{q^2\big\}^{-1}\bigg]\bigg) \bigg(2 \psi
-\frac{c}{2 \sqrt{6} \sqrt{H_{0}^2 \Psi^{2 q+2}}}\bigg) \Psi^{2
q+2}+\bigg(\bigg[\bigg\{ \bigg(12 H_{0}^2
\\
\nonumber&\times&\Psi^{2 q+2} \psi-\sqrt{6} c \bigg(\sqrt{H_{0}^2
\Psi^{2 q+2}}-H_{0} \Psi^{\sigma}\bigg)\bigg)\sigma^2 \Psi^{-2
q}\bigg\}\big\{q^2\big\}^{-1}\bigg]-12 n^2\bigg) \\
\nonumber&\times&\sqrt{6} c \sigma^2 \bigg(12
n^2-\bigg[\bigg\{\Psi^{-2 q}\sigma^2 \bigg(12 H_{0}^2 \Psi^{2 q+2}-
\bigg(\sqrt{H_{0}^2 \Psi^{2 q+2}}-H_{0} \Psi^{\sigma}\bigg)\\
\nonumber&\times&\sqrt{6}
c\bigg)\bigg\}\big\{q^2\big\}^{-1}\bigg]\bigg) \sqrt{H_{0}^2 \Psi^{2
q+2}}\bigg)\bigg\}\bigg\{2 q^4 \bigg(\bigg[\bigg\{ c \sigma^2
\bigg(\sqrt{H_{0}^2 \Psi^{2 q+2}}-H_{0} \Psi^{\sigma}\bigg)\\
\nonumber&\times&\sqrt{6}\Psi^{-2
q}\bigg\}\big\{q^2\big\}^{-1}\bigg]+12 n^2\bigg)^2
\bigg(\bigg[\bigg\{\bigg(12 H_{0}^2 \Psi^{2 q+2}-\sqrt{6} c
\bigg(\sqrt{H_{0}^2 \Psi^{2 q+2}}\\
\nonumber&-&H_{0} \Psi^{\sigma}\bigg)\bigg)\Psi^{-2
q}\sigma^2\bigg\}\big\{q^2\big\}^{-1}\bigg]-12
n^2\bigg)^2\bigg\}^{-1}\bigg].
\end{eqnarray}
Figure \textbf{4} shows how the freezing region is calculated for
different values of $\psi$ and $n$, where
$\omega_{D}<0,~\omega^{\prime}_{D}<0$. This indicates an
acceleration in cosmic expansion at higher rates in this context.
\begin{figure}
\epsfig{file=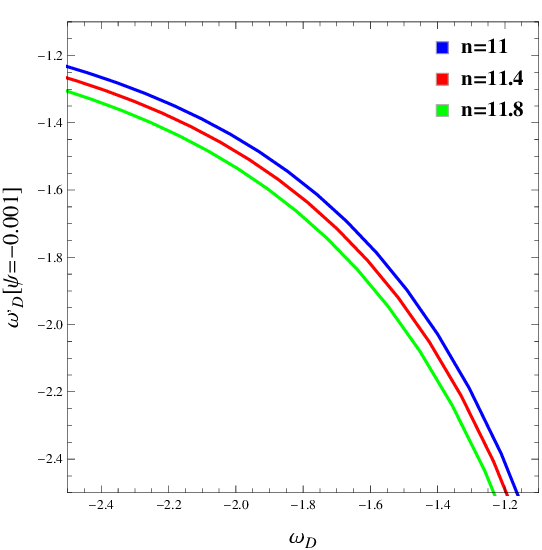,width=.4\linewidth}
\epsfig{file=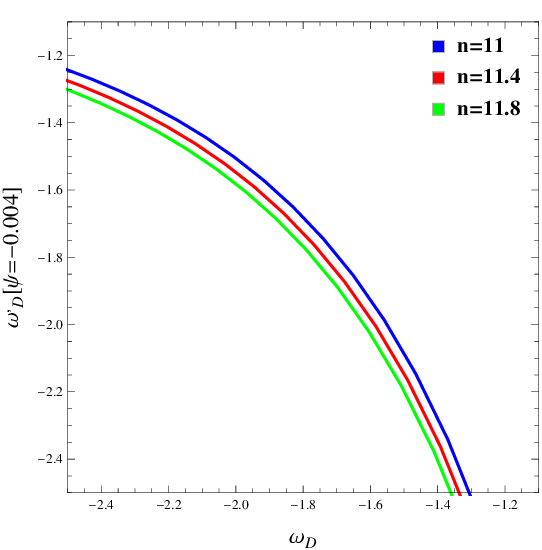,width=.4\linewidth}\center
\epsfig{file=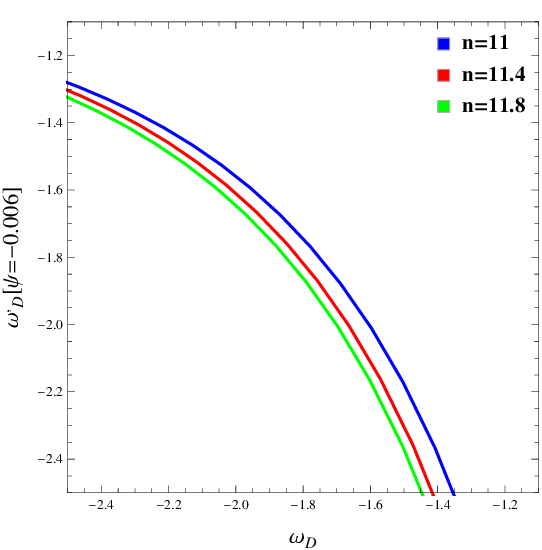,width=.4\linewidth} \caption{Graphs of
$\omega^{\prime}_{D}$ versus $\omega_{D}$.}
\end{figure}

\subsection{The $(r-s)$-Plane}

One way to explore the the universe's dynamics from a cosmological
viewpoint is through statefinder $(r,s)$ analysis \cite{46}.
Understanding various DE models require this essential approach.
Trajectories are classified as part of the quintessence and phantom
phases if they exist in the region $(r<1;s>0)$, while the Chaplygin
gas models manifested when $(r>1;s<0)$. The flat universe is
characterized by these specific parameters
\begin{equation}\nonumber
r=\frac{\dddot{a}}{aH^{3}}, \quad s=\frac{r-1}{3(q-\frac{1}{2})}.
\end{equation}
The cosmos consists of two distinct parts of the EoS parameters,
$\omega_{D}$ and $\omega_{m}$, representing exotic energy and
ordinary matter, respectively. The values $(r,s)$ are defined as
\begin{equation}\nonumber
r=1+\frac{9\omega_{D}}{2}\Omega_{D}(1+\omega_{D})
-\frac{3\omega^{\prime}_{D}}{2H}\Omega_{D},\quad
s=1+\omega_{D}-\frac{\omega^{\prime}_{D}}{3\omega_{D}H}.
\end{equation}
These parameters for the NADE $f(\mathcal{Q})$ gravity turn out to
be
\begin{eqnarray}\nonumber
r&=& \bigg[\bigg\{-24 \eta ^4 n^2 \bigg(3 c^2 \bigg(6 H^2 \bigg(3
\sqrt{6}-9 \mathcal{Q}^{3/2}-4 \sqrt{\mathcal{Q}}\bigg)+\bigg(9
\sqrt{6} \mathcal{Q}^2-15
\sqrt{\mathcal{Q}}\\
\nonumber&+&4 \sqrt{6} \mathcal{Q}\bigg)2 H -4 \mathcal{Q}^{3/2} -9
\mathcal{Q}^{5/2}+2 \sqrt{6} \mathcal{Q}\bigg)+2 c \bigg(-H\bigg(9
\sqrt{6} \mathcal{Q}^{3/2} (2 \psi +1)
\\
\nonumber&-&72 \psi+16 \sqrt{6} \sqrt{\mathcal{Q}}\bigg)+9
\mathcal{Q}^2 (2 \psi +1)-9 \sqrt{6} \sqrt{\mathcal{Q}} \psi +16
\mathcal{Q}\bigg)\mathcal{Q}+12 \sqrt{6} \mathcal{Q}^2 \psi
\\
\nonumber&-&16\mathcal{Q}^{5/2}\bigg)+\eta ^6 \bigg(c^3 \bigg(-12
H^3 \bigg(9 \sqrt{6} \mathcal{Q}^{3/2}+4 \sqrt{6}
\sqrt{\mathcal{Q}}-18\bigg)-\bigg(5
\sqrt{6}\sqrt{\mathcal{Q}}\\
\nonumber&-&8 \mathcal{Q}-18 \mathcal{Q}^2\bigg)18 H^2 -6 H \bigg(9
\sqrt{6} \mathcal{Q}^{3/2}+4 \sqrt{6}
\sqrt{\mathcal{Q}}-12\bigg)\mathcal{Q}-3 \sqrt{6} \mathcal{Q}^{3/2}
\\
\nonumber&+&18 \mathcal{Q}^3+8 \mathcal{Q}^2\bigg)-2 c^2\mathcal{Q}
\bigg(6 H^2 \bigg(9 \mathcal{Q}^{3/2} (2 \psi+1)+16
\sqrt{\mathcal{Q}}-12 \sqrt{6} \psi \bigg) -2H\\
\nonumber&\times&\bigg(16 \sqrt{6} \mathcal{Q}-54 \sqrt{\mathcal{Q}}
\psi+9 \sqrt{6} \mathcal{Q}^2 (2 \psi +1)\bigg)+9 \mathcal{Q}^{5/2}
(2 \psi +1)+16 \mathcal{Q}^{3/2}-6 \\
\nonumber&\times&\sqrt{6} \mathcal{Q}\psi \bigg)+4 c \mathcal{Q}^2
\bigg(36 H \psi-8 \sqrt{6} H \sqrt{\mathcal{Q}}-3 \sqrt{6}
\sqrt{\mathcal{Q}}\psi +8 \mathcal{Q}\bigg)+72 \mathcal{Q}^{9/2}
\psi ^2\bigg)\\
\nonumber&-&144 \eta ^2 n^4 \bigg(3 c \bigg(2 H \bigg(9 \sqrt{6}
\mathcal{Q}^{3/2}+4 \sqrt{6} \sqrt{\mathcal{Q}}-18\bigg)-18
\mathcal{Q}^2-8 \mathcal{Q}+5 \sqrt{6}
\sqrt{\mathcal{Q}}\bigg) \\
\nonumber&+&18 \mathcal{Q}^{\frac{5}{2}} (2 \psi +1)-24 \sqrt{6}
\mathcal{Q} \psi+32 \mathcal{Q}^{\frac{3}{2}}\bigg)-3456 n^6 \bigg(3
\sqrt{6}-9
\mathcal{Q}^{\frac{3}{2}}-4 \sqrt{\mathcal{Q}}\bigg)\bigg\}\\
\nonumber&\times&\bigg\{ \sqrt{\mathcal{Q}} \big(c\eta ^2
\big(\sqrt{\mathcal{Q}}-\sqrt{6} H\big)+12 n^2\big) \big(\eta ^2
\big(\sqrt{6}
 c H-c \sqrt{\mathcal{Q}}+ \mathcal{Q}\big)-12 n^2\big)^2\bigg\}^{-1}\bigg],\\
\nonumber s&=& \bigg[\bigg\{-24 \eta ^4 n^2 \bigg(c^2 \bigg(18 H^2
\bigg(\sqrt{6}-3 \mathcal{Q}^{\frac{3}{2}}\bigg)+6 H \bigg(3
\sqrt{6} \mathcal{Q}^2-5
\sqrt{\mathcal{Q}}\bigg)+\sqrt{6} \mathcal{Q}\\
\nonumber&-&9 \mathcal{Q}^{\frac{5}{2}}\bigg)+6 c \bigg(-H
\bigg(\sqrt{6} \mathcal{Q}^{\frac{5}{2}} (2 \psi +1)-8 \mathcal{Q}
\psi \bigg)-\sqrt{6} \mathcal{Q}^{\frac{3}{2}} \psi +\mathcal{Q}^3
(2 \psi
+1)\bigg)\\
\nonumber&+&4 \sqrt{6} \mathcal{Q}^2 \psi \bigg)+\eta ^6 \bigg(c^3
\bigg(-36 H^3 \bigg(\sqrt{6} \mathcal{Q}^{3/2}-2\bigg)-6 H^2 \bigg(5
\sqrt{6}
\sqrt{\mathcal{Q}}-18 \mathcal{Q}^2\bigg)\\
\nonumber&-&6 H \bigg(3 \sqrt{6} \mathcal{Q}^{\frac{3}{2}}-4\bigg)
\mathcal{Q}-\sqrt{6} \mathcal{Q}^{\frac{3}{2}}+6
\mathcal{Q}^3\bigg)-2 c^2 \mathcal{Q} \bigg(
\bigg(\mathcal{Q}^{\frac{3}{2}} (6 \psi +3)-4
\sqrt{6} \psi \bigg)\\
\nonumber&\times&6 H^2-6 H \bigg(\sqrt{6} \mathcal{Q}^2 (2 \psi
+1)-6 \sqrt{\mathcal{Q}} \psi \bigg)+\mathcal{Q}^{\frac{5}{2}} (6
\psi +3)-2 \sqrt{6} \mathcal{Q} \psi \bigg)+ c
\\
\nonumber&\times&\bigg(48 H\mathcal{Q}^2 \psi -4 \sqrt{6}
\mathcal{Q}^{5/2} \psi \bigg)+24 \mathcal{Q}^{9/2} \psi ^2\bigg)-144
\bigg(\bigg(18 H
\bigg(\sqrt{6} \mathcal{Q}^{3/2}-2\bigg)\\
\nonumber&-&18 \mathcal{Q}^2+5 \sqrt{6} \sqrt{\mathcal{Q}}\bigg)c+6
\mathcal{Q}^{5/2} (2 \psi +1)-8 \sqrt{6} \mathcal{Q} \psi \bigg)\eta
^2 n^4-3456 n^6 \bigg(\sqrt{6}\\
\nonumber&-&3 \mathcal{Q}^{3/2}\bigg)\bigg\} \bigg\{6
\mathcal{Q}^{3/2}\bigg(c \eta ^2 \bigg(\sqrt{\mathcal{Q}}-\sqrt{6}
H\bigg)+12 n^2\bigg) \bigg(\eta ^2 \bigg(\sqrt{6} c H-c
\sqrt{\mathcal{Q}}\\
\nonumber&+&2 \mathcal{Q}\bigg)-12 n^2\bigg)\bigg(\eta ^2
\bigg(\sqrt{6} c H-c \sqrt{\mathcal{Q}}+2 \mathcal{Q} \psi \bigg)-12
n^2\bigg)\bigg\}^{-1}\bigg],
\end{eqnarray}
while in the context of $z$, we get
\begin{eqnarray}\nonumber r&=&
\bigg[\bigg\{-3456 \sqrt{6} n^6 \bigg(3-54 \bigg(H_{0}^2 \Psi^{2
q+2}\bigg)^{3/2} -4 \sqrt{H_{0}^2 \Psi^{2
q+2}}\bigg)-\bigg[\bigg\{432 n^4 \sigma^2  \\
\nonumber&\times&\Psi^{-2 q}\bigg(64 \sqrt{6} \bigg(H_{0}^2 \Psi^{2
q+2}\bigg)^{3/2}+216 \sqrt{6} (2 \psi +1) \bigg(H_{0}^2 \Psi^{2
q+2}\bigg)^{5/2}+\bigg(\bigg(54\\
\nonumber&\times&\bigg(H_{0}^2 \Psi^{2 q+2}\bigg)^{3/2}+4
\sqrt{H_{0}^2 \Psi^{2 q+2}}-3\bigg) 2 H_{0} \Psi^{\sigma}-8 H_{0}^2
\Psi^{2 q+2}-108 H_{0}^4 \Psi^{4 q+4}\\
\nonumber&+&5 \sqrt{H_{0}^2 \Psi^{2 q+2}}\bigg)6 c -48 \sqrt{6}
H_{0}^2 \Psi^{2 q+2} \psi
\bigg)\bigg\}\big\{q^2\big\}^{-1}\bigg]+\bigg[\bigg\{\sigma^6
\Psi^{-6 q} \bigg(93312 \sqrt{6} \psi ^2 \\
\nonumber&\times&\bigg(H_{0}^2 \Psi^{2 q+2}\bigg)^{9/2}+6 c^3
\bigg(-18 \bigg(H_{0}^2 \Psi^{2 q+2}\bigg)^{3/2}+48 H_{0}^4 \Psi^{4
q+4}+648 H_{0}^6 \Psi^{6 q+6}\\
\nonumber&-&3 H_{0}^2 \bigg(30 \sqrt{H_{0}^2 \Psi^{2 q+2}}-48
H_{0}^2 \Psi^{2 q+2}-648 H_{0}^4 \Psi^{4 q+4}\bigg)\Psi^{2 q+2}-72
H_{0}^3 \Psi^{3 q+3} \\
\nonumber&\times&\bigg(27 \bigg(H_{0}^2 \Psi^{2 q+2}\bigg)^{3/2}-1+2
\sqrt{H_{0}^2 \Psi^{2 q+2}}\bigg) -12 H_{0}^3 \Psi^{3 q+3} \bigg(54
\bigg(H_{0}^2 \Psi^{2 q+2}\bigg)^{3/2} \\
\nonumber&+&4 \sqrt{H_{0}^2 \Psi^{2 q+2}}-3\bigg)\bigg)+864H_{0}^4c
\bigg(6 H_{0} \psi \Psi^{\sigma}-8 H_{0} \sqrt{H_{0}^2 \Psi^{2 q+2}}
\Psi^{\sigma}+8 H_{0}^2 \Psi^{2 q+2}\\
\nonumber&-&3 \sqrt{H_{0}^2 \Psi^{2 q+2}} \psi \bigg)\Psi^{4 q+4}
-144 \sqrt{6} c^2 H_{0}^2 \Psi^{2 q+2} \bigg(8 \bigg(H_{0}^2 \Psi^{2
q+2}\bigg)^{3/2} +27 (2 \psi +1)
\\
\nonumber&\times&\bigg(H_{0}^2 \Psi^{2 q+2}\bigg)^{5/2}-3 H_{0}^2
\Psi^{2 q+2} \psi -H_{0} \Psi^{\sigma} \bigg(16 H_{0}^2 \Psi^{2
q+2}+54 H_{0}^4 (2 \psi +1) \Psi^{4 q+4} \\
\nonumber&-&9 \sqrt{H_{0}^2 \Psi^{2 q+2}} \psi \bigg)+H_{0}^2
\Psi^{2 q+2} \bigg(27 (2 \psi +1) \bigg(H_{0}^2 \Psi^{2
q+2}\bigg)^{3/2}+8 \sqrt{H_{0}^2 \Psi^{2 q+2}}\\
\nonumber&-&6 \psi \bigg)\bigg)
\bigg)\bigg\}\big\{q^6\big\}^{-1}\bigg]-\bigg[\bigg\{144 n^2
\sigma^4 \Psi^{-4 q} \bigg(-96 \sqrt{6} \bigg(H_{0}^2 \Psi^{2
q+2}\bigg)^{5/2}+3 \sqrt{6} c^2\\
\nonumber&\times&\bigg(-4 \bigg(H_{0}^2 \Psi^{2 q+2}\bigg)^{3/2}-54
\bigg(H_{0}^2 \Psi^{2 q+2}\bigg)^{5/2}+2 H_{0}^2 \Psi^{2 q+2}+H_{0}
\Psi^{\sigma} \bigg(8 H_{0}^2 \Psi^{2 q+2}\\
\nonumber&+&108 H_{0}^4 \Psi^{4 q+4}-5 \sqrt{H_{0}^2 \Psi^{2 q+2}}
\bigg)+ \bigg(3-54 \bigg(H_{0}^2 \Psi^{2 q+2}\bigg)^{3/2}-4
\sqrt{H_{0}^2 \Psi^{2 q+2}}\bigg)\\
\nonumber&\times&H_{0}^2 \Psi^{2 q+2}\bigg)+72 \sqrt{6} H_{0}^4
\Psi^{4 q+4} \psi +2 c H_{0}^2 \Psi^{2 q+2} \bigg(96 H_{0}^2 \Psi^{2
q+2}- \bigg(324 (2 \psi +1)\\
\nonumber&\times&\bigg(H_{0}^2 \Psi^{2 q+2} \bigg)^{3/2}+96
\sqrt{H_{0}^2 \Psi^{2 q+2}}-72 \psi \bigg) H_{0}\Psi^{\sigma}+324
H_{0}^4 (2 \psi +1) \Psi^{4 q+4}\\
\nonumber&\times&\Psi^{-2 q}\bigg\}\big\{q^2\big\}^{-1}\bigg]-54
\sqrt{H_{0}^2 \Psi^{2 q+2}} \psi
\bigg)\bigg)\bigg\}\big\{q^4\big\}^{-1}\bigg]\bigg\}\bigg\{
\sqrt{H_{0}^2 \Psi^{2 q+2}} \bigg(\bigg[\bigg\{\sqrt{6} c \sigma^2
\\
\nonumber&\times&\bigg(\sqrt{H_{0}^2 \Psi^{2 q+2}}-H_{0}
\Psi^{\sigma}\bigg)+12 n^2\bigg)8 \sqrt{6}
\bigg(\bigg[\bigg\{\sigma^2 \Psi^{-2 q} \bigg(12 H_{0}^2 \Psi^{2
q+2}-\sqrt{6} c
\\
\nonumber&\times&\bigg(\sqrt{H_{0}^2 \Psi^{2 q+2}}-H_{0}
\Psi^{\sigma}\bigg)\bigg)\bigg\}\big\{q^2\big\}^{-1}\bigg]-12 n^2\bigg)^2
\bigg\}^{-1}\bigg]. \\
\nonumber s&=& \bigg[\bigg\{-576 \sqrt{6} n^6 \bigg(1-18
\bigg(H_{0}^2 \Psi^{2 q+2}\bigg)^{\frac{3}{2}}\bigg)-\bigg[\bigg\{24
n^4
\sigma^2 \Psi^{-2 q} \bigg(216 \sqrt{6} (2 \psi +1)\\
\nonumber&\times&\bigg(H_{0}^2 \Psi^{2 q+2}\bigg)^{\frac{5}{2}}+c
\bigg(18 H_{0} \bigg(36 \bigg(H_{0}^2 \Psi^{2
q+2}\bigg)^{\frac{3}{2}}-2\bigg)
\Psi^{\sigma}-648 H_{0}^4 \Psi^{4 q+4}+30\\
\nonumber&\times&\sqrt{H_{0}^2 \Psi^{2 q+2}}\bigg)-48 \sqrt{6}
H_{0}^2 \Psi^{2 q+2} \psi
\bigg)\bigg\}\big\{q^2\big\}^{-1}\bigg]-\bigg[\bigg\{24 n^2 \sigma^4
\Psi^{-4 q} \bigg(\sqrt{6} c^2 H_{0} \\
\nonumber&\times&\bigg(108 H_{0}^3 \Psi^{3 q+3}+5\bigg) \bigg(H_{0}
\Psi^{\sigma}-\sqrt{H_{0}^2 \Psi^{2 q+2}}\bigg) \Psi^{\sigma}+24
\sqrt{6} H_{0}^4 \psi  \Psi^{4 q+4}+c \\
\nonumber&\times&\bigg(216 H_{0}^6 \Psi^{6 q+6} (2 \psi +1)-36 \psi
\bigg(H_{0}^2 \Psi^{2 q+2}\bigg)^{\frac{3}{2}}-H_{0}
\Psi^{\sigma} \bigg(216 \bigg(H_{0}^2 \Psi^{2 q+2}\bigg)^{\frac{5}{2}}\\
\nonumber&\times&(2 \psi +1)-48 H_{0}^2 \Psi^{2 q+2} \psi
\bigg)\bigg)\bigg)\bigg\}\big\{q^4\big\}^{-1}\bigg]+\bigg[\bigg\{\sigma^6
\bigg(84 \sqrt{6} \psi ^2 \bigg(H_{0}^2 \Psi^{2
q+2}\bigg)^{\frac{9}{2}}\\
\nonumber&+&c^3 \bigg(216 H_{0}^6 \Psi^{6 q+6}-6 \bigg(H_{0}^2
\Psi^{2 q+2}\bigg)^{\frac{3}{2}}- \bigg(30 \sqrt{H_{0}^2 \Psi^{2
q+2}}-648
H_{0}^4 \Psi^{4 q+4}\bigg)\\
\nonumber&\times&H_{0}^2 \Psi^{2 q+2}-6 H_{0}^3 \Psi^{3 q+3} \big(36
\big(H_{0}^2 \Psi^{2 q+2}\big)^{\frac{3}{2}}-2\big)- \bigg(108
\big(H_{0}^2 \Psi^{2 q+2}\big)^{\frac{3}{2}}-4\bigg)\\
\nonumber&\times&6 H_{0}^3 \Psi^{3 q+3}\bigg)+144 c \bigg(2 H_{0}^5
\Psi^{5 q+5}-\bigg(H_{0}^2 \Psi^{2 q+2}\bigg)^{\frac{5}{2}}\bigg)
\psi -24
\sqrt{6} c^2 H_{0}^2 \Psi^{2 q+2} \\
\nonumber&\times&\bigg(3 (6 \psi +3) \bigg(H_{0}^2 \Psi^{2
q+2}\bigg)^{5/2}-H_{0}^2 \Psi^{2 q+2} \psi -3 H_{0} \Psi^{\sigma}
\bigg(6 H_{0}^4 \Psi^{4 q+4} (2 \psi +1)\\
\nonumber&-&\sqrt{H_{0}^2 \Psi^{2 q+2}} \psi \bigg)+H_{0}^2 \Psi^{2
q+2} \bigg(3 \bigg(H_{0}^2 \Psi^{2 q+2}\bigg)^{\frac{3}{2}} (6 \psi
+3)-2
\psi \bigg)\bigg)\bigg)\Psi^{-6q}\bigg\}\\
\nonumber&\times&\big\{q^6\big\}^{-1}\bigg]\bigg\}\bigg\{6 \sqrt{6}
\big(H_{0}^2 \Psi^{2 q+2}\big)^{\frac{3}{2}}
\bigg(\bigg[\bigg\{\sqrt{6} c \sigma^2 \bigg(\sqrt{H_{0}^2 \Psi^{2
q+2}}-H_{0} \Psi^{\sigma}\bigg)
\Psi^{-2 q}\bigg\}\\
\nonumber&\times&\big\{q^2\big\}^{-1}\bigg]+12 n^2\bigg)
\bigg(\bigg[\bigg\{\sigma^2  \bigg(12 H_{0}^2 \Psi^{2 q+2}-\sqrt{6}
c \bigg(\sqrt{H_{0}^2 \Psi^{2 q+2}}-H_{0}
\Psi^{\sigma}\bigg)\bigg)\\
\nonumber&\times&\Psi^{-2 q}\bigg\}\big\{q^2\big\}^{-1}\bigg]-12
n^2\bigg) \bigg(\bigg[\bigg\{\sigma^2 \Psi^{-2 q} \bigg(12 H_{0}^2
\Psi^{2 q+2} \psi -\sqrt{6} c \bigg(\sqrt{H_{0}^2 \Psi^{2
q+2}}\\
\nonumber&-&H_{0}
\Psi^{\sigma}\bigg)\bigg)\bigg\}\big\{q^2\big\}^{-1}\bigg]-12
n^2\bigg)\bigg\}^{-1}\bigg].
\end{eqnarray}
For all values of $n$ and $\psi$, Figure \textbf{5} depicts the
behavior of the $(r-s)$-plane as the Chaplygin gas model.
\begin{figure}
\epsfig{file=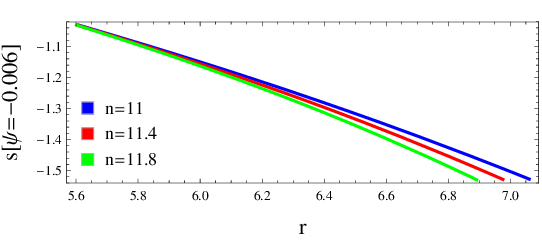,width=0.49\linewidth}
\epsfig{file=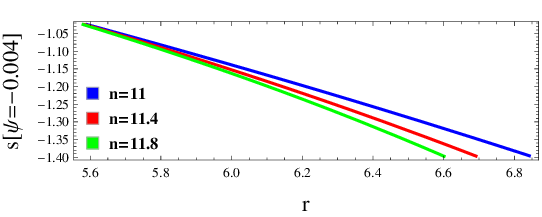,width=0.53\linewidth}\center
\epsfig{file=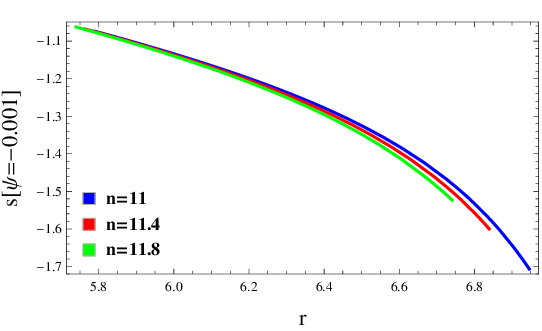,width=.5\linewidth} \caption{Graphs of $s$
versus $r$.}
\end{figure}

\subsection{The Squared Speed of Sound Parameter}

The squared speed of sound parameter can be expressed as
\begin{equation}\label{28}
\nu_{s}^{2}=\frac{P_{GGDE}}{\rho'_{GGDE}}\omega'_{GGDE}
+\omega_{GGDE}.
\end{equation}
The signature of $\nu_{s}^{2}$ is essential in analyzing the
stability of the reconstructed NADE model. The presence of a
positive $\nu_{s}^{2}$ indicates stability, while a negative
$\nu_{s}^{2}$ denotes instability in the model. The corresponding
$\nu_{s}^{2}$ is given as
\begin{eqnarray}\nonumber
\nu_{s}^{2}&=&\bigg[\bigg\{\bigg(c \eta ^2 \bigg(Q (2 e-Q)+6 H^2
Q-H\bigg)-12 n^2 Q^{3/2}\bigg) \bigg(-\bigg[\bigg\{
\bigg(\eta ^2 \bigg(\sqrt{6} c H\\
\nonumber&-&c \sqrt{Q}+2 Q \psi \bigg)-12 n^2\bigg)\eta
^2\bigg\}\bigg\{\bigg(c \eta ^2 \big(\sqrt{Q}-\sqrt{6}
H\big)+12 n^2\bigg) \bigg(\eta ^2 \bigg(\big(\sqrt{Q}\\
\nonumber&-&\sqrt{6} H\big)c-2 Q\bigg)+12
n^2\bigg)\bigg\}^{-1}\bigg]+\bigg[\bigg\{c  \sqrt{Q}
\bigg(\eta ^2 \bigg(\sqrt{6} c H-c \sqrt{Q}+2 Q \psi \bigg)\\
\nonumber&-&12 n^2\bigg)\eta ^4\bigg\}\bigg\{2 \bigg(c \eta ^2
\big(\sqrt{Q}-\sqrt{6} H\big)+12 n^2\bigg)^2 \bigg(\eta ^2 \bigg(c
\big(\sqrt{Q}-\sqrt{6} H\big)-2\\
\nonumber&\times& Q\bigg)+12
n^2\bigg)\bigg\}^{-1}\bigg]+\bigg[\bigg\{\eta ^4 Q \bigg(\frac{c}{2
\sqrt{Q}}-2\bigg) \bigg(\eta ^2 \bigg(\sqrt{6} c
H-c \sqrt{Q}+2 Q \psi \bigg)\\
\nonumber&-&12 n^2\bigg)\bigg\}\bigg\{\bigg(c \eta ^2
\bigg(\sqrt{Q}-\sqrt{6} H\bigg)+12 n^2\bigg) \bigg(\eta ^2 \bigg(c
\bigg(\sqrt{Q}-\sqrt{6} H\bigg)-2 Q\bigg)\\
\nonumber&+&12 n^2\bigg)^2\bigg\}^{-1}\bigg]-\bigg[\bigg\{\eta ^4 Q
\bigg(2 \psi -\frac{c}{2 \sqrt{Q}}\bigg)\bigg\}\bigg\{\bigg(c \eta
^2 \bigg(\sqrt{Q}-\sqrt{6} H\bigg)+12 n^2\bigg)\\
\nonumber&\times&\bigg(\eta ^2 \bigg(c \big(\sqrt{Q}-\sqrt{6}
H\big)-2 Q\bigg)+12
n^2\bigg)\bigg\}^{-1}\bigg]\bigg)\bigg\}\bigg\{\bigg[\bigg\{c \big(2
\eta ^2 Q^{3/2}\big)\bigg\}\bigg\{
\sqrt{Q}\\
\nonumber&\times&4\bigg\}^{-1}\bigg]\bigg\}^{-1}\bigg]-\bigg[\bigg\{\eta
^2 Q \bigg(\eta ^2 \bigg(\sqrt{6} c H-c \sqrt{Q}+2 Q \psi \bigg)-12
n^2\bigg)\bigg\}\bigg\{\bigg(c \eta ^2 \\
\nonumber&\times&\bigg(\sqrt{Q}-\sqrt{6} H\bigg)+12 n^2\bigg)
\bigg(\eta ^2 \bigg(c \bigg(\sqrt{Q}-\sqrt{6} H\bigg)-2 Q\bigg)+12
n^2\bigg)\bigg\}^{-1}\bigg],
\end{eqnarray}
while in the context of the $z$, it can be observed that
\begin{eqnarray}\nonumber
\nu_{s}^{2}&=&\bigg[\bigg\{\bigg[\bigg\{\Psi^{-4 q} \bigg(-72
\sqrt{6} n^2 \bigg(H_{0}^2 \Psi^{2 q+2}\bigg)^{3/2}-\bigg[\bigg\{c
H_{0}\bigg(12 H_{0}^2 \Psi^{3 q+3}+1\bigg)\sigma^2
\\
\nonumber&\times&\Psi^{1-q}\bigg\}\big\{q^2\big\}^{-1}\bigg]\bigg)
\bigg(-2 q^2 \bigg(\bigg[\bigg\{\sqrt{6} c \sigma^2
\bigg(\sqrt{H_{0}^2 \Psi^{2 q+2}}-H_{0} \Psi^{\sigma}\bigg) \Psi^{-2
q}\bigg\}\\
\nonumber&\times&\big\{q^2\big\}^{-1}\bigg]+12 n^2\bigg) \bigg(12
n^2-\bigg[\bigg\{\sigma^2 \Psi^{-2 q} \bigg(12 H_{0}^2 \Psi^{2 q+2}-
\bigg(\sqrt{H_{0}^2 \Psi^{2 q+2}}\\
\nonumber&-&H_{0} \Psi^{\sigma}\bigg)\sqrt{6}
c\bigg)\bigg\}\big\{q^2\big\}^{-1}\bigg]\bigg)
\bigg(\bigg[\bigg\{\sigma^2 \Psi^{-2 q} \bigg(12 H_{0}^2 \Psi^{2
q+2} \psi - \bigg(\sqrt{H_{0}^2 \Psi^{2 q+2}}\\
\nonumber&-&H_{0} \Psi^{\sigma}\bigg)\sqrt{6}
c\bigg)\bigg\}\big\{q^2\big\}^{-1}\bigg]-12 n^2\bigg) \Psi^{2 q}+12
H_{0}^2 \sigma^2 \bigg(\bigg[\bigg\{\sqrt{6} c
\bigg(\sqrt{H_{0}^2 \Psi^{2 q+2}}\\
\nonumber&-&H_{0} \Psi^{\sigma}\bigg) \sigma^2\Psi^{-2
q}\bigg\}\big\{q^2\big\}^{-1}\bigg]+12 n^2\bigg)
\bigg(\bigg[\bigg\{\sigma^2 \Psi^{-2 q} \bigg(12 H_{0}^2 \Psi^{2
q+2} \psi -\sqrt{6} c\\
\nonumber&\times&\bigg(\sqrt{H_{0}^2 \Psi^{2 q+2}}-H_{0}
\Psi^{\sigma}\bigg)\bigg)\bigg\}\big\{q^2\big\}^{-1}\bigg]-12
n^2\bigg) \bigg(\bigg[\bigg\{2 \sqrt{6} \sqrt{H_{0}^2
\Psi^{2 q+2}}\bigg\}^{-1}\\
\nonumber&\times&\big\{c\big\}\bigg]-2\bigg) \Psi^{2 q+2}-12 H_{0}^2
\sigma^2 \bigg(\bigg[\bigg\{\sqrt{6} c \sigma^2 \bigg(\sqrt{H_{0}^2
\Psi^{2 q+2}}-H_{0} \Psi^{\sigma}\bigg) \Psi^{-2
q}\bigg\}\\
\nonumber&\times&\big\{q^2\big\}^{-1}\bigg]+12 n^2\bigg) \bigg(12
n^2-\bigg[\bigg\{\sigma^2 \Psi^{-2 q} \bigg(12 H_{0}^2 \Psi^{2
q+2}-\sqrt{6} c \big(\sqrt{H_{0}^2 \Psi^{2 q+2}}\\
\nonumber&-&H_{0}
\Psi^{\sigma}\big)\bigg)\bigg\}\big\{q^2\big\}^{-1}\bigg]\bigg)
\bigg(2 \psi -\bigg[\big\{c\big\}\bigg\{2 \sqrt{6} \sqrt{H_{0}^2
\Psi^{2 q+2}}\bigg\}^{-1}\bigg]\bigg) \Psi^{2 q+2}+\sqrt{6}\\
\nonumber&\times&c \sigma^2 \bigg(\bigg[\bigg\{\sigma^2 \Psi^{-2 q}
\bigg( H_{0}^2 \Psi^{2 q+2} \psi -\sqrt{6} c \bigg(\sqrt{H_{0}^2
\Psi^{2 q+2}}-H_{0}
\Psi^{\sigma}\bigg)\bigg)\bigg\}\big\{q^2\big\}^{-1}\bigg]\\
\nonumber&-&12 n^2\bigg) \bigg(12 n^2-\bigg[\bigg\{\sigma^2 \Psi^{-2
q} \bigg(H_{0}^2 \Psi^{2 q+2}-\sqrt{6} c \bigg(\sqrt{H_{0}^2 \Psi^{2
q+2}}-H_{0}
\Psi^{\sigma}\bigg)\bigg)\bigg\}\\
\nonumber&\times&\big\{q^2\big\}^{-1}\bigg]\bigg) \sqrt{H_{0}^2
\Psi^{2 q+2}}\bigg)\bigg\}\bigg\{c \bigg(\bigg[\bigg\{\sigma^2
\Psi^{-2 q} \bigg( H_{0}^2 \Psi^{2 q+2}-\sqrt{6} c
\bigg(\sqrt{H_{0}^2 \Psi^{2 q+2}}\\
\nonumber&-&H_{0}\Psi^{\sigma}\bigg)\bigg)\bigg\}\big\{q^2\big\}^{-1}\bigg]-12
n^2\bigg)^2\bigg\}^{-1}\bigg]-\bigg[\bigg\{H_{0}^4 \sigma^2 \Psi^4
\bigg(\bigg[\bigg\{\sqrt{6} c \bigg(\sqrt{H_{0}^2
\Psi^{2 q+2}}\\
\nonumber&-&H_{0} \Psi^{\sigma}\bigg) \sigma^2\Psi^{-2
q}\bigg\}\big\{q^2\big\}^{-1}\bigg]+12 n^2\bigg)
\bigg(\bigg[\bigg\{\sigma^2  \bigg(12 H_{0}^2 \Psi^{2 q+2} \psi
-\bigg(\sqrt{H_{0}^2 \Psi^{2 q+2}}\\
\nonumber&-&H_{0} \Psi^{\sigma}\bigg)\sqrt{6} c \bigg)\Psi^{-2
q}\bigg\}\big\{q^2\big\}^{-1}\bigg]-12 n^2\bigg)\bigg\}\bigg\{12
n^2-\bigg[\bigg\{\sigma^2 \Psi^{-2 q} \bigg(12 H_{0}^2 \Psi^{2
q+2}\\
\nonumber&-&\sqrt{6} c \bigg(\sqrt{H_{0}^2 \Psi^{2 q+2}}-H_{0}
\Psi^{\sigma}\bigg)\bigg)\bigg\}\big\{q^2\big\}^{-1}\bigg]\bigg\}\bigg\{6
H_{0}^2 q^2 \Psi^2 \bigg(\bigg[\bigg\{\sigma^2 \bigg(\sqrt{H_{0}^2
\Psi^{2 q+2}}\\
\label{1aaa}&-&H_{0} \Psi^{\sigma}\bigg) \sqrt{6} c\Psi^{-2
q}\bigg\}\big\{q^2\big\}^{-1}\bigg]+12
n^2\bigg)^2\bigg\}^{-1}\bigg].
\end{eqnarray}
Several studies have investigated this aspect for different DE
models. For instance, Setare \cite{2abc} examined the interacting
HDE model with the Chaplygin gas and found that both models exhibit
instability. Kim et al. \cite{3abc} showed that the NADE model is
always negative, indicating its instability. Figure \textbf{6} shows
that the NADE $f(\mathcal{Q})$ model is unstable for all values of
$n$ and $\psi$, as the $\nu_{s}^{2}$ remains negative throughout the
evolution of the universe. This aligns with the previous studies,
showing that the $f(\mathcal{Q})$ model faces similar instability
challenges in the literature.
\begin{figure}
\epsfig{file=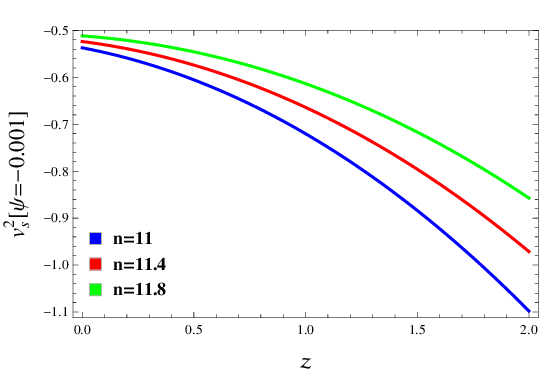,width=0.5\linewidth}
\epsfig{file=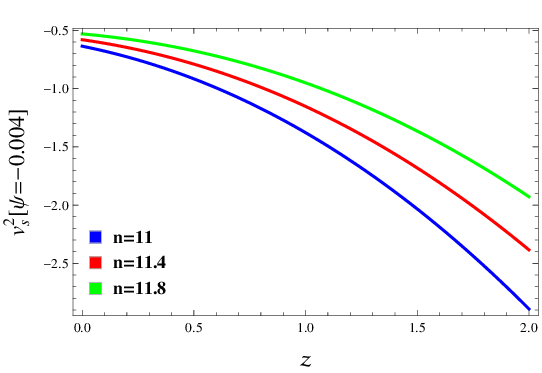,width=0.5\linewidth}\center
\epsfig{file=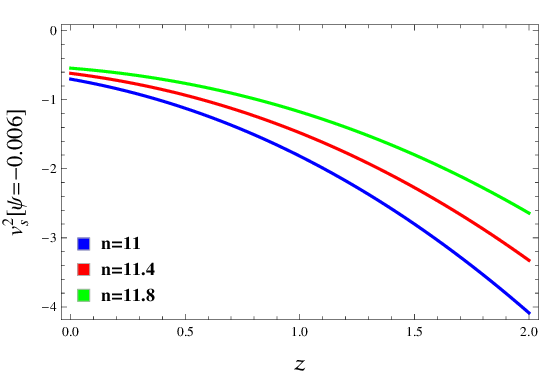,width=0.5\linewidth}\caption{Graphs of
$\nu_{s}^{2}$ versus $z$.}
\end{figure}

\section{Conclusions}

In this paper, we have examined the NADE model in the
$f(\mathcal{Q})$ gravity. Initially, we have utilized the
correspondence scheme to reconstruct the NADE $f(\mathcal{Q})$
gravity model. We have applied the FRW model with a power-law
expression for the scale factor in the interacting scenario. We have
assumed that the densities of NADE and $f(\mathcal{Q})$ gravity are
equal to find NADE $f(\mathcal{Q})$ gravity model. We have
graphically analyzed the behavior of NADE model for three distinct
values of $n=11,11.4,11.8$. We have analyzed the EoS,
$(\omega_{D}-\omega^{\prime}_{D})$ and $(r-s)$ planes. The
$\nu_{s}^{2}$ is employed to analyze the stability of the
interacting NADE $f(\mathcal{Q})$ gravity model. The key findings
are outlined as follows.
\begin{itemize}
\item
The NADE $f(\mathcal{Q})$ gravity model shows an increasing pattern
for both $z$ and $\mathcal{Q}$ with selected values of $n$,
indicating the realistic nature of the reconstructed model (Figure
\textbf{1}).
\item
The energy density demonstrates a positive trend, while the pressure
shows negative behavior for all values of $n$. These observations
align with the typical features of DE (Figure \textbf{2}).
\item
In the later stages of evolution, it is observed that $\omega_{D}$
behaves as the quintessence-like characteristic for power-law form
using various values of $n$ and $\psi$ (Figure \textbf{3}). It is
noted that the rate of evolution of the energy density could be
sufficiently slow at present time resolving the coincidence problem.
\item
The evolutionary pattern of the
($\omega_{D}$-$\omega^{\prime}_{D}$)-plane shows the region where
freezing occurs for chossen values of $\psi$ and $n$ (Figure
\textbf{4}). This indicates that the NADE $f(\mathcal{Q})$ gravity
suggests a more rapid expansion of the universe.
\item
The $(r-s)$-plane depicts the Chaplygin gas model for various values
of $\psi$ and $n$ (Figure \textbf{5}).
\item
We have determined that the $\nu_{s}^{2}$ is negative, indicating
instability for selected values of $\psi$ and $n$ in the NADE
$f(\mathcal{Q})$ gravity (Figure \textbf{6}).
\end{itemize}

Jawad et al. \cite{47} examined the NADE model in the
$f(\mathcal{G})$ gravity ($\mathcal{G}$ is the Gauss-Bonnet
invariant) to analyze the expansion of the universe and assessed the
stability of the model. They found that NADE $f(\mathcal{G})$ model
demonstrated a graphically decreasing behavior. The reconstructed
NADE $f(\mathcal{Q})$ model shows a graphical increasing trend. Both
models are unstable as the universe evolves and both show a
quintessence region for acceleration observed through the EoS. The
key concept in theoretical progressions involves higher-order
gravitational theories that incorporate anti-gravity phenomena due
to higher-order curvature terms. It is important to mention here
that the $f(\mathcal{Q})$ theory is better suited for addressing the
above mentioned problem as compared to $f(\mathcal{G})$ since the
field equations of $f(\mathcal{Q})$ gravity are second order,
whereas the field equations of $f(\mathcal{G})$ gravity are fourth
order.\\\\
\textbf{Data Availability Statement:} No data was used for the
research described in this paper.

\end{document}